# Giant Interfacial Thermal Resistance Arising From Materials With Mismatched Phonon Structures

Yan Yin[1,2]*, Busheng Wang[1], Jin Chen[1], Jiyong Yao[3], Wenlong Wang[1,2,4], Li Wang[1], Huaizhou Zhao[1,5]*, Wuming Liu[1,2,4]*, Xuedong Bai[1,2,4]*

[1]Beijing National Laboratory for Condensed Matter Physics and Institute of Physics, Chinese Academy of Sciences, Beijing 100190, China.

[2]School of Physical Sciences, University of Chinese Academy of Sciences, Beijing 100190, China.

[3]Center for Crystal Research and Development, Key Laboratory of Functional Crystals and Laser Technology, Technical Institute of Physics and Chemistry, Chinese Academy of Sciences, Beijing 100190, China.

[4]Songshan Lake Materials Laboratory, Dongguan, Guangdong 523808, China.

[5]The Yangtze River Delta Physics Research Center, Liyang, Jiangsu, China.

*Correspondence to: yan.yin@iphy.ac.cn; hzhao@iphy.ac.cn; wmliu@iphy.ac.cn; xdbai@iphy.ac.cn.

**Abstract:**

Previous researches only reported very small interfacial thermal resistances at room temperature due to limitations in sample combinations and methods. Taking cognizance of the importance of mismatched phonon structures, we report values up to $2\times10^{-4}$ $W^{-1}m^2K$, thousand times larger than highest values reported to date. This enables substantial tuning of the thermal conductivity in composites, and doesn't constrain other characteristics. Our findings inspire new design strategies, for heat control in integrated circuits and thermoelectric composites, that harness thermal transport at interfaces.

*Introduction* - Interfacial thermal resistance (ITR), also known as thermal boundary resistance or Kapitza resistance, is the resistance to thermal flow at an interface that arises due to differences in the materials. It is intrinsically directional and matters in almost all heat control situations. Research on ITR started in the 1940s and the field of low-temperature physics, when Kapitza observed the thermal resistance at interfaces between copper and liquid helium [1,2]. ITRs between solids at low temperature were measured with steady-state methods in the 1970s and 1980s [3,4]. For example, the ITR of sapphire-indium is about $40\times10^{-4}$ $W^{-1}m^2K$ at 1K temperature [5]. Due to a $T^{-3}$ temperature relationship, ITR was expected to be very small at room temperature. Complex transient methods were developed to measure ITRs at room temperature, of thin metal films on dielectrics [6-8], thin dielectric films on crystal-structure-matched dielectric substrates [9], thin metal films on metals [10,11], and thin-films of superlattice [12]. As expected, these reports fed back very small and a narrow range of ITRs ($10^{-7}$-$10^{-10}$ $W^{-1}m^2K$), even smaller than the radiation limit, which is the theoretical minimum possible resistance for transport via the elastic process of phonons at a given interface [6]. With such small values, ITRs would exert only limited influence on and offer very limited tuning abilities for physical processes.

Most theoretical approaches used to evaluate ITR are based on simple models of classical acoustic wave impedances [6,7] or center on a single parameter, namely, the Debye temperature [6,8]. These methods work well for low-temperature conditions, under which phonons condense in low-energy acoustic branches. However, they do not consider any contribution from the specific conditions related to the matching or mismatching of the phonon structures, which can be very complex at room temperature when all phonon modes are populated. In addition, the ITR was always assumed to be much smaller than the contact resistance due to imperfection or the roughness of the interfaces between solids. This preset assumption greatly limited the explored combinations between solid-solid samples to metal-metal, dielectric-metal, and crystal-structure-matched dielectrics.

Here, we used both steady-state and transient methods to study the ITRs between macroscopically thick layers of different semiconductors or semiconductor-metal combinations. Some ITRs can be as large as the contact resistance due to the roughness of the interfaces, and dominate in the transient laser flash method. Only for combinations with highly mismatched phonon structures can the ITR be as large as $2\times10^{-4}$ $W^{-1}m^2K$, which is three orders of magnitude larger than the largest value reported to date at room temperature [8]. We also offer a simple theoretical model based on the detailed phonon density of state (DOS) structures of the materials to approximate the phonon heat transport mismatch factor at an interface. The behavior of the calculated mismatch factors across different material combinations is consistent with the behavior of the measured resistances. This consistency supports the important role of phonon structure mismatch in determining the ITR at room temperature. Our results show that pure ITR-based control of the effective thermal conductivities in composites can offer a range of at least five orders of magnitude, only 1-2 orders of magnitude fewer than those spanned by the thermal conductivities of all known bulk materials. In addition, this phenomenon offers a feasible means of independently controlling the thermal transport properties of a two-element composite with no direct limitations on its other properties. This is because ITR is associated with the mismatching of the phonon structures between the materials, and each individual material is still in macroscopic size and maintains their bulk properties.

*Quantitative results.* - Our experimental results show that ITR can be significantly (three orders of magnitude) larger than previously reported values, and the thermal conductivities of two-element composites can be made to span at least five orders of magnitude (from $10^{-1}$ to $10^3$ $Wm^{-1}K^{-1}$) purely by adjusting the ITR. The samples used for the heat flux method were tungsten disulfide ($WS_2$), silicon (Si, Si-0.58, Si-0.35), $BaGa_4Se_7$ (BGS), $Mn_{0.75}Cu_{0.25}Sb_2Se_4$ (MCSS), $Bi_2Te_3$, and copper (Cu). Table I lists all the contact thermal resistances ($R_{contact}$), ITRs ($R_{interfacial}$) and interfacial thermal conductances (ITCs) between the two layers. The contact thermal resistance is obtained by calculating the difference between the stacking thermal resistance ($R_{combined}$) of two layers and the sum of the resistances of two individual layers. The ITR is obtained by calculating the difference between the contact resistance of given samples with a heterogeneous interface and the contact resistance of the same samples but with a homogenous interface. For example, the contact thermal resistance of $WS_2$ sample 1 and Si sample 1 is 0.23 K/W, the contact thermal resistance of $WS_2$ sample 1 and $WS_2$ sample 2 is 0.20 K/W, and the contact thermal resistance of Si sample 1 and Si sample 2 is 0.049 K/W. The average of the contact thermal resistances of the $WS_2$-$WS_2$ interface and the Si-Si interface is 0.12 K/W. This value is regarded as the homogenous contact resistance, which is purely due to interface roughness or imperfection, of the heterogeneous $WS_2$-Si interface. Then the difference between 0.23 K/W and 0.12 K/W is the ITR, which is the contribution due to the heterogeneous material mismatch, of the $WS_2$-Si interface.

From Table I, we can see that the ITCs of the different combinations can range from $4.6\times10^3$ $Wm^{-2}K^{-1}$ to larger than $160\times10^3$ $Wm^{-2}K^{-1}$. The minimum value in ITCs corresponds to $2\times10^{-4}$ $W^{-1}m^2K$ in ITR, which is three orders of magnitude larger than the largest value reported to date at room temperature [8]. Moreover, one can always tune an ITR to be nonexistent or negligible by erasing an interface or using identical materials, and this effectively pushes the upper limit on the ITC to infinity.

Next, we calculated the effective thermal conductivities corresponding to the ITCs in Table I under the assumption of one interface every 0.1 mm in thickness. In the "stacking of the ITR" section of the Supplementary Materials, we proved that the ITRs in Table I are the same if multiple alternating layers are used and the thickness between layers is at the order of 0.1mm. The calculated values are plotted as red stars in Fig. 1, together with the thermal conductivities of most bulk materials. The thermal conductivities of all materials only span approximately six to seven orders of magnitude (from $10^{-2}$ or $10^{-3}$ to $10^3$ $Wm^{-1}k^{-1}$). Figure 1 shows that our results - purely considering the effect of the ITC - correspond to a range of at least five orders of magnitude. It is very likely that the ITRs are still stackable using an alternating-layer thickness smaller than 0.1 mm, as the heat-transport-weighted phonon free length for silicon is reported to be on the order of 1 μm [13]; in this case the effective thermal conductivities attainable may be smaller, potentially comparable to or surpassing those of the best thermal insulation materials. More importantly, this ability to tune the thermal conductivity is directly associated with the matching or mismatching of the phonon structures between the materials, and has no direct relationship with the properties of the individual materials. At the thickness of 0.1 mm, each individual material will maintain its bulk material properties, unlike superlattice structures. This capability offers a path for controlling the thermal transport properties of a two-element composite independently of its other major properties.

*AB composites for large resistances.* - It appears that AB-type material combinations such as those discussed next yield large ITRs, tested by both steady-state and transient methods. Type A materials have reasonable hardness (the maximum phonon energy is approximately 500 $cm^{-1}$), high crystal symmetry (simple and certain slopes in phonon dispersion curves), and a large phonon gap or quasigap. Type B materials are very soft (the maximum phonon energy is below 300 $cm^{-1}$) and have low crystal symmetry (complex and mostly flat phonon structures).

Here, we used $WS_2$ and Si as type A materials and BGS and MCSS as type B materials, and tested the interface properties of all possible combinations. Table II lists the measured contact thermal resistances ($R_{contact}$) and ITR ($R_{interfacial}$) for combinations of these four materials. There are two subtables, one for $R_{contact}$ and another for $R_{interfacial}$. In each subtable, the results from the steady-state heat flux method are presented in the upper right triangle and provide the quantitative values, whereas the results from the transient laser flash method are presented in the bottom left triangle and offer a better data contrast between heterogeneous and homogenous interfaces.

The data obtained by steady-state method clearly show larger (more than double) interfacial resistances for AB-type combinations than for materials of the same type. As explained in the Supplementary Materials, the heat pulse delay – an indicator of the transient thermal resistance - at an interface is used to investigate the contact resistance in the laser flash method. The contact resistances of the homogenous interfaces were considered to be the baseline values and were subtracted from the heterogeneous interface values to obtain the ITR values shown in the second subtable of Table II. Other than elevated values for heterogeneous combinations involving Si due to the large differences in thermal diffusivities between layers, the results from the laser flash method show similar behavior as those from the heat flux method. By exploring all combinations of these four materials of two material types with both steady-state and transient methods, we have shown that this kind of AB combination indeed results in a large ITR in both steady and transient states.

Our hypothesis is that the mismatching of the phonon structures between AB materials gives rise to the large ITR. The calculated phonon dispersion curves and phonon DOSs of $WS_2$, Si, BGS, and $MnSb_2Se_4$ are shown in Fig. 2. The phonon structure of MCSS is approximately same as that of $MnSb_2Se_4$. We can see that $WS_2$ and Si (with a phonon quasigap, a wide middle area with a low phonon DOS weight) fit the description for type A and that BGS and MCSS fit the description for type B. The phonon structures of A- and B-type materials are visually very different. Beside the visual differences, next we will use a simple theoretical method to approximately evaluate the mismatch and show that the mismatches of AB combinations are indeed significantly larger than those of same-type combinations.

We employ a very simple calculation method that considers only the mismatching of the energy and DOS. We estimate the contribution of the phonon structure to the thermal transport to be as: $K = n(E) \cdot E \cdot v_g$. $K$ is the differential thermal transport contribution from phonons with a particular energy $E$, $n(E)$ is the phonon number at $E$ and includes the phonon DOS function $g(E)$, and $v_g$ is the phonon group velocity at $E$. $v_g$ is considered to have an average effective value for all values of $E$. $K$ was calculated using the phonon DOS in Fig. 2 and a temperature of 80°C (our measurement temperature) and was then normalized to satisfy $\int K \cdot dE = 1$. The resulting $K$ functions are plotted in Fig. S2.

We define a phonon heat transport mismatch factor, as an approximate indicator for ITR, in the form 1-$f_{overlap}$, where $f_{overlap}$ is the area of overlap between the $K$ functions of any two materials. The mismatch factors were calculated from Fig. S2 for all combinations of $WS_2$, Si, BGS, and MCSS; the results are listed in Table III. Mismatch factors of AB combinations range from 0.53 to 0.77; mismatch factors of same type combinations range from 0 to 0.52. These values support the conclusion that AB-type materials possess noticeable larger mismatches in phonon heat transmission than do same-type material combinations. This finding is consistent with the experimental results. Overall, our theoretical calculations support the hypothesis that the mismatching of the phonon structures between AB materials gives rise to the large ITR.

*Potential applications.* - ITR offers a feasible means of independently controlling the thermal transport properties of a two-element composite with no direct limitations on its other properties. Extreme-anisotropic thermal conductivity composite is used as an example to explain this concept. Another example, thermoelectric two-element composite, is discussed in the Supplementary Materials.

In vertically stacked integrated circuits, high-temperature or high-power sources need to be cooled and isolated at the same time, because other heat-sensitive devices are nearby and the space is very limited. A thermal material or design with extreme-anisotropic thermal conductivity will be very useful, can dissipate heat fast along one direction to a heat sink and isolate heat from other devices in another direction.

We can expect a composite made with multiple alternating layers of Si or silicon carbide and another much softer semiconductor/material, and their ITR is reasonable large. Let's assume the ITC is $4.6\times10^3$ Wm$^{-2}$K$^{-1}$, same as the smallest value in Table I. The ITR is intrinsically directional, and the thermal conductivity in the out-of-plane direction will be 0.46 Wm$^{-1}$K$^{-1}$, for 0.1-mm-thick layers. The in-of-plane thermal conductivity of silicon carbide is 500 Wm$^{-1}$K$^{-1}$, same as the bulk material. The anisotropic ratio is 1100, which is about 5 times of the 212 of graphite - a high anisotropic thermal conductivity material. If ITR is still fully functional for 10-μm-thick layers, then the anisotropic ratio can be above 10000.

Even better, this heat control strategy can be integrated into circuits design. For example, the silicon carbide layers can be used for functional layers and/or middle layers, while the other phonon-mismatched material can be used in middle layers. It will enable heat cooling along in-plane direction (even in the heat isolation middle layers) and isolation along out-plane direction, thus achieve directional heat control in integrated circuits.

*Summary.* - We investigated the ITRs between macroscopically-thick layers of different semiconductors or semiconductor-metal composites at room temperature, using both steady-state and transient experiments as well as theoretical calculations. At room temperature, ITR can be as large as $2\times10^{-4}$ W$^{-1}$m$^2$K. By controlling the ITR, the effective thermal conductivities in composites can be made to span at least five orders of magnitude. The mismatching of phonon structures plays an important role in determining ITR. In the steady state situation, ITR can be as large as or even larger than the contact resistance due to surface roughness; in the transient laser flash measurement, the impact of contact resistance due to surface roughness is very small or ignorable. Tailoring the ITR through phonon mismatch offers a practical way to tune the thermal transport properties of composites, and doesn't directly constrain any of the other characteristics of the bulk materials involved. We present new design strategies, for thermoelectric two-element composites and directional heat control in vertically stacked integrated circuits, as examples of the potential applications of this phenomenon.

**Acknowledgments:** Yan Yin thanks Prof. WeiYa Zhou, Prof. Kuijuan Jin, Prof. Sheng Meng and Prof. Yulong Liu for their helps and discussions. Yan Yin also thanks Yu Zhao for assisting in preparing samples, and the staff at Centre Testing International (CTI) for assisting with measurements.

**Funding:** This work was supported by the National Natural Science Foundation of China (Grant Nos. 11574388, 21773303, 11474337, 51572287, and U1601213).

**Supplementary Materials:**

Materials and methods

Stacking of the ITR

Factors affecting the ITR

Thermoelectric two-element composites

Figures S1-S3

Table S1-S2

An information pdf file about Long Win's LW-9389 instrument

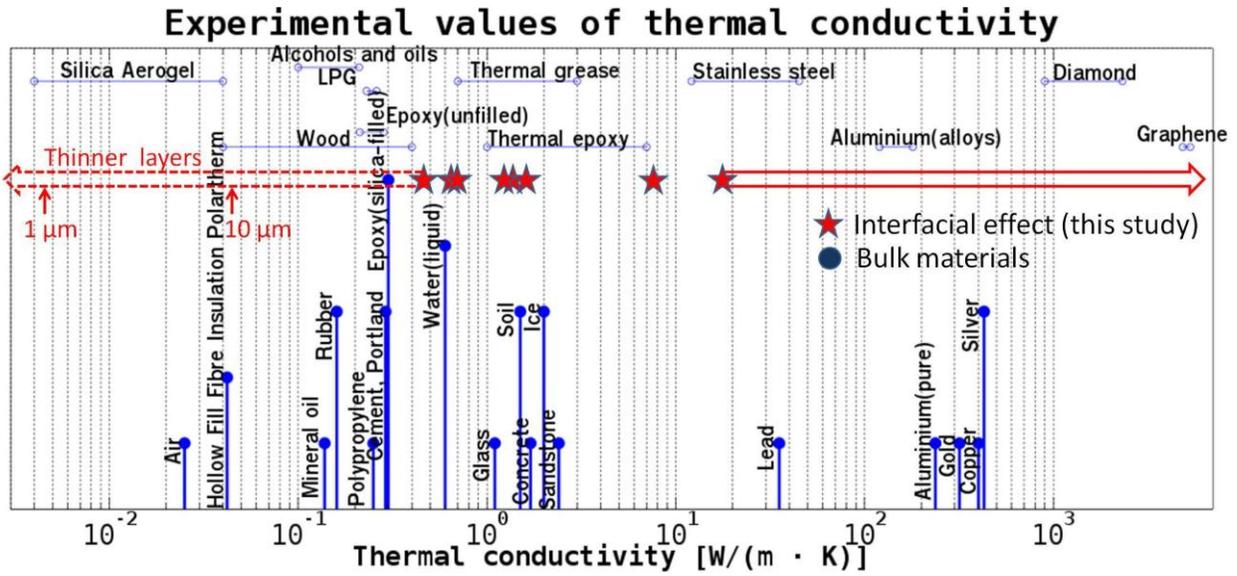

FIG. 1. Thermal conductivity of a variety of bulk materials and the interfacial effect studied in this work. Red stars represent purely interfacial effective thermal conductivities calculated using the ITC values from Table I and the assumption of 0.1-mm-thick alternating layers. The solid red arrow pointing to the right represents the covering range by interfacial effect but out of our measurement range. The dash red arrow pointing to the left represents the possible covering range by interfacial effect, if the ITRs are still stackable with thinner layers. The two vertical red arrows mark the corresponding low limit positions for 1- or 10-μm-thick alternating layers. Conductivity values for bulk materials and the background figure are taken from https://en.wikipedia.org/wiki/Thermal_conductivity.

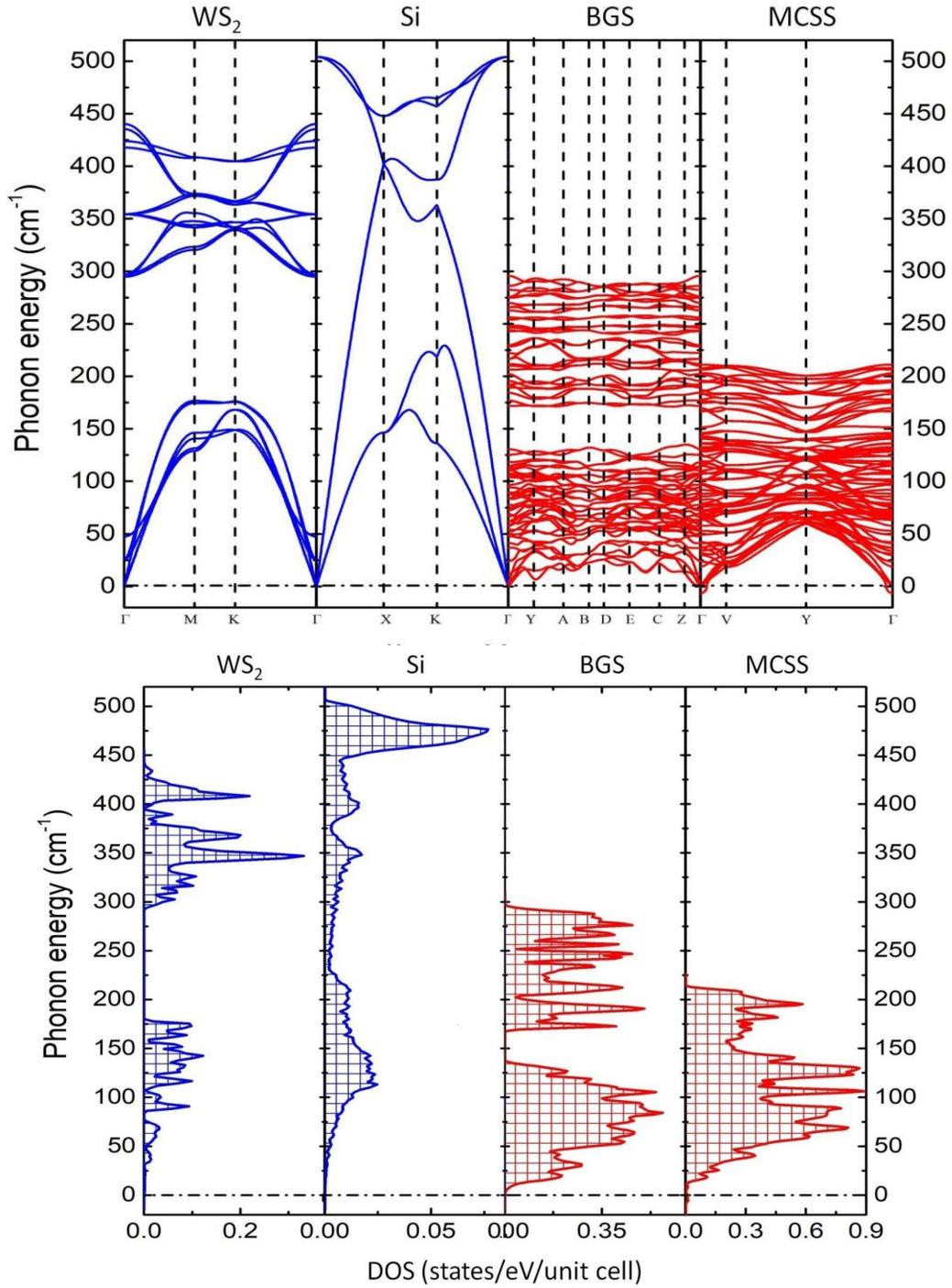

FIG. 2. Calculated phonon dispersion curves and phonon DOS values of $WS_2$, Si, $BaGa_4Se_7$ (BGS), and $MnSb_2Se_4$. Blue color is used for A type, and red color is used for B type materials. The phonon structure of MCSS is approximately same as that of $MnSb_2Se_4$.

TABLE I. Contact thermal resistance (R$_{contact}$), ITR (R$_{interfacial}$) and ITC between two layers for all layer combinations tested.

| Composition of each two-layer composite | | Individual-layer resistance, R$_{individual}$ (K/W) | | Combined resistance of layers, R$_{combined}$ (K/W) | Contact thermal resistance, R$_{contact}$ (K/W) | Interfacial thermal resistance, R$_{interfacial}$ (K/W) | Interfacial area (cm$^2$) | Interfacial thermal conductance (10$^3$ Wm$^{-2}$K$^{-1}$) |
|---|---|---|---|---|---|---|---|---|
| Layer I | Layer II | Layer I | Layer II | | | | | |
| WS$_2$ | WS$_2$ | 1.26 | 1.51 | 2.97 | 0.20 | 0 | 6.25 | - |
| Si | Si | 0.009 | 0.009 | 0.067 | 0.049 | 0 | 6.25 | - |
| BGS | BGS | 1.74 | 1.79 | 3.83 | 0.30 | 0 | 6.25 | - |
| MCSS | MCSS | 1.53 | 1.72 | 3.53 | 0.28 | 0 | 6.25 | - |
| WS$_2$ | Si | 1.26 | 0.009 | 1.50 | 0.23 | 0.11 | 6.25 | 15 |
| WS$_2$ | BGS | 1.26 | 1.76 | 3.52 | 0.50 | 0.25 | 6.25 | 6.4 |
| WS$_2$ | MCSS | 1.26 | 1.67 | 3.52 | 0.59 | 0.35 | 6.25 | 4.6 |
| Si | BGS | 0.01 | 1.74 | 2.15 | 0.40 | 0.23 | 6.25 | 7.0 |
| Si | MCSS | 0.01 | 1.53 | 1.93 | 0.39 | 0.23 | 6.25 | 7.0 |
| BGS | MCSS | 1.74 | 1.53 | 3.56 | 0.29 | 0.00 | 6.25 | ≥160 |
| Bi$_2$Te$_3$ | Bi$_2$Te$_3$ | 1.39 | 1.52 | 3.03 | 0.11 | 0 | 6.76 | - |
| Si-0.35 | Si-0.35 | 0.006 | 0.006 | 0.047 | 0.035 | 0 | 6.76 | - |
| Bi$_2$Te$_3$ | Si | 1.52 | 0.009 | 1.73 | 0.20 | 0.12 | 6.76 | 12 |
| Bi$_2$Te$_3$ | Si-0.58 | 1.40 | 0.014 | 1.62 | 0.20 | 0.13 | 6.76 | 11 |
| Bi$_2$Te$_3$ | Si-0.35 | 1.40 | 0.006 | 1.61 | 0.20 | 0.13 | 6.76 | 11 |
| Cu | Cu | 0.002 | 0.002 | 0.079 | 0.075 | 0 | 6.76 | - |
| Cu | Si-0.35 | 0.002 | 0.006 | 0.087 | 0.079 | 0.02 | 6.76 | 74 |

TABLE II. Contact thermal resistance ($R_{contact}$) and ITR ($R_{interfacial}$) obtained at interfaces between $WS_2$, Si, $BaGa_4Se_7$ (BGS), and $Mn_{0.75}Cu_{0.25}Sb_2Se_4$ (MCSS), as measured by the steady-state heat flux method and the transient laser flash method. Results from the steady-state heat flux method are presented in the upper right triangles. Values of the heat-pulse delay Δt due to the interface – an indicator of the thermal resistance – as measured with the transient laser flash method are presented in the bottom left triangles for direct comparison with the values obtained using the steady-state method.

| Transient method | Steady-state method | A type | | B type | | $R_{contact}$ (K/W) | |
|---|---|---|---|---|---|---|---|
| | | $WS_2$ | Si | BGS | MCSS | | |
| | | 0.20 | 0.23 | 0.50 | 0.59 | $WS_2$ | A type |
| A type | $WS_2$ | 0.00 | | 0.05 | 0.40 | 0.39 | Si |
| | Si | 0.22 | 0.01 | | 0.30 | 0.29 | BGS | B type |
| B type | BGS | 0.32 | 0.31 | 0.03 | | 0.28 | MCSS |
| | MCSS | 0.52 | 0.55 | 0.06 | 0.06 | | Heat flux method |
| $\Delta t_{contact}$ (s) | | $WS_2$ | Si | BGS | MCSS | Laser flash method | |
| | | A type | | B type | | | |

| Transient method | Steady-state method | A type | | B type | | $R_{interfacial}$ (K/W) | |
|---|---|---|---|---|---|---|---|
| | | $WS_2$ | Si | BGS | MCSS | | |
| | | 0 | 0.11 | 0.25 | 0.35 | $WS_2$ | A type |
| A type | $WS_2$ | 0 | | 0 | 0.23 | 0.23 | Si |
| | Si | 0.22 | 0 | | 0 | 0.00 | BGS | B type |
| B type | BGS | 0.30 | 0.29 | 0 | | 0 | MCSS |
| | MCSS | 0.49 | 0.51 | 0.02 | 0 | | Heat flux method |
| $\Delta t_{interfacial}$ (s) | | $WS_2$ | Si | BGS | MCSS | Laser flash method | |
| | | A type | | B type | | | |

TABLE III. Calculated phonon heat transport mismatch factors between layers of $WS_2$, Si, $BaGa_4Se_7$ (BGS), and $Mn_{0.75}Cu_{0.25}Sb_2Se_4$ (MCSS). The phonon structure of MCSS is approximately considered to be same as that of $MnSb_2Se_4$.

| $WS_2$ | Si | BGS | MCSS | Mismatch factor |
|---|---|---|---|---|
| 0 | 0.52 | 0.77 | 0.62 | $WS_2$ |
|  | 0 | 0.64 | 0.53 | Si |
|  |  | 0 | 0.34 | BGS |
|  |  |  | 0 | MCSS |